\documentclass[prl,twocolumn,showpacs,superscriptaddress]{revtex4}
\usepackage{graphicx,amsfonts,amsmath,amssymb} 
\usepackage{times,txfonts}

\def\kbar{\protect\@kbar}
\def\@kbar{%
\relax \bgroup
\def\@tempa{\hbox{\raise.73\ht0
\hbox to0pt{\kern.25\wd0\vrule width.5\wd0
height.1pt depth.1pt\hss}\box0}}%
\mathchoice{\setbox0\hbox{$\displaystyle k$}\@tempa}%
{\setbox0\hbox{$\textstyle k$}\@tempa}%
{\setbox0\hbox{$\scriptstyle k$}\@tempa}%
{\setbox0\hbox{$\scriptscriptstyle k$}\@tempa}%
\egroup}

\begin{document}

\title{Gravity-Sensitive Quantum Dynamics in Cold Atoms}
\author{Z.Y. Ma}
\affiliation{Clarendon Laboratory, Department of Physics,
University of Oxford, Oxford OX1 3PU, United Kingdom}
\author{M.B. d'Arcy}
\affiliation{Atomic Physics Division, National Institute of
Standards and Technology, Gaithersburg, Maryland 20899-8424, USA}
\author{S.A. Gardiner}
\affiliation{JILA, University of Colorado and National Institute
of Standards and Technology, Boulder, Colorado 80309-0440, USA}

\date{\today}

\begin{abstract}
We subject a falling cloud of cold cesium atoms to periodic kicks
from a sinusoidal potential created by a vertical standing wave of
off-resonant laser light. By controllably accelerating the
potential, we show quantum accelerator mode dynamics to be highly
sensitive to the effective gravitational acceleration when this is
close to specific, resonant values. This quantum sensitivity to a
control parameter is reminiscent of that associated with classical
chaos, and promises techniques for precision measurement.
\end{abstract}


\maketitle

The identification and observation of signatures of chaos in
quantum dynamics is the goal of considerable current effort. Much
of this work centers on the theoretical definition and
characterization of energy spectra \cite{Haake2001}, or such
quantities as the Loschmidt echo \cite{Cucchietti2002} and
fidelity \cite{Cerruti2002}, which essentially develop the idea
that sensitivity of a wavefunction's evolution to small variations
in a system's Hamiltonian be used as a definition of quantum
instability \cite{Cucchietti2002,Cerruti2002,Haake2001,Peres1993}.
Such quantities could be observed experimentally but require some
interpretation to highlight the way in which their nature betokens
stability or chaos. An attractive alternative would be the
observation of different motional regimes. This is more in sympathy
with the techniques and philosophy used to identify classical
chaos, and is the approach used here.

In certain systems the decay of the overlap of two initially
identical wavefunctions evolving under slightly differing
Hamiltonians can be expressed in the long time limit as the sum of
two decay predictions, governed by Fermi's Golden Rule and the
classical Lyapunov exponent \cite{Cucchietti2002}. The decay rate
serves as a quantum signature of instability, which can be
compared with that of the corresponding classical system. Such
sensitivity can be probed by interferometric techniques
\cite{Gardiner1997,Schlunk2003a}. In the quantum-mechanical system
presented here, the classical limit of which is chaotic, extreme
sensitivity of the
qualitative nature of the motional dynamics to a control parameter
is directly observable. It is manifested by the effect on
quantum accelerator mode (QAM) dynamics
\cite{Schlunk2003a,dArcy2001a,Oberthaler1999,Godun2000,Schlunk2003b}
of small variations in the effective value of gravity in the
$\delta$-kicked accelerator \cite{dArcy2001a}, an extension of the
paradigmatic $\delta$-kicked rotor \cite{Fishman1993}. The QAM
observed in this atom optical realization
\cite{Schlunk2003a,dArcy2001a,Oberthaler1999,Godun2000,Schlunk2003b},
are characterized by a momentum transfer, linear with kick number,
to a substantial fraction (up to $\sim 20$\%) of the initial cloud
of atoms. This is due to a resonant rephasing effect, dependent on
the time-interval between kicks, for certain initial wavefunctions
\cite{Godun2000,Fishman2002}. The sensitivity in the dynamics we
observe also promises the capability of precisely calibrating a
relationship between the local gravitational acceleration and
$h/m$, where $m$ is the atomic mass, and we describe how our
observations constitute a preliminary feasibility-demonstration of
such a measurement.

The Hamiltonian of the $\delta$-kicked accelerator, realized using
a magneto-optic trap (MOT) of laser-cooled atoms that are then
released and subjected to pulses from a standing wave of
off-resonant light, is
\begin{equation}
\hat{H}=\frac{\hat{p}^2}{2m}+mg\hat{z}-\hbar\phi_d[1+\cos(G\hat{z})]\sum\limits_{n=-\infty}
^{+\infty}\delta(t-nT),
\end{equation}
where $\hat{z}$ is the position, $\hat{p}$ the momentum, $m$ the
particle mass, $t$ the time, $T$ the pulse period, $\hbar\phi_d$
quantifies the strength of the kicking potential,
$G=2\pi/\lambda_{\mbox{\scriptsize spat}}$, and
$\lambda_{\mbox{\scriptsize spat}}$ is the spatial period of the
standing wave applied to the atoms. The quantity $g$ is normally
the gravitational acceleration. However, by `accelerating' the
standing wave, it is possible to effectively modify $g$. We have
previously used this technique to counteract gravity and regain
kicked rotor dynamics \cite{dArcy2001a,dArcy2001b}.

In an innovative analysis by Fishman, Guarneri, and Rebuzzini
(FGR) \cite{Fishman2002}, the fact that QAM are observed only when
$T$ approaches $\ell T_{1/2}=\ell 2\pi m /\hbar G^{2}$, where
$\ell\in\mathbb{Z}^{+}$ and $T_{1/2}$ is the half-Talbot time
\cite{Godun2000}, is exploited to yield a dramatically simplified
picture of QAM dynamics. In a frame accelerating with $g$, the
linear potential is removed to leave a spatially periodic
Hamiltonian. The quasimomentum $\beta$ is then conserved,
i.e., if
a momentum state $|p\rangle=|(k+\beta)\hbar G\rangle$,
where $k\in \mathbb{Z}$ and $\beta\in[0,1)$, `ladders' of
momentum states of different $\beta$ evolve {\it independently}.
The resulting $n$-dependent 
and $\beta$-specific kick-to-kick time evolution operator is
\begin{equation}
\begin{split}
\hat{F}_{n}(\beta) =& \exp(-i \{ \hat{\rho}+\mbox{sgn}(\epsilon) [
\pi\ell + \kbar \beta
- \gamma(n-1/2) ] \}^{2} /2\epsilon )\\ & \times \exp (
i\tilde{k}\cos\hat{\chi}/|\epsilon| ),
\end{split}
\label{Eq:TEvolve}
\end{equation}
where $\tilde{k}=|\epsilon|\phi_{d}$, $\kbar=2\pi T/T_{1/2}$, and
$\gamma=gGT^{2}$. We have introduced a smallness parameter,
$\epsilon=2\pi(T/T_{1/2}-\ell)$, to 
quantify the closeness of $T$ to
$\ell T_{1/2}$ and the dynamical variables are now an
angle $\hat{\chi}=G\hat{z}$ and a discrete conjugate momentum
$\hat{\rho}=\hat{p}|\epsilon|/\hbar G$, such that
$[\hat{\chi},\hat{\rho}]=i|\epsilon|$. If one constructs a
kick-to-kick Heisenberg map corresponding to Eq.\
(\ref{Eq:TEvolve}) for the dynamical variables, then in the limit
$\epsilon\rightarrow 0$, the commutator vanishes along with the
uncertainty principle, and the operators can be replaced by their
mean values. Thus
\begin{subequations}
\begin{gather}
\tilde{\rho}_{n+1}=\tilde{\rho}_n-\tilde{k}\sin(\chi_n)-\mbox{sgn}(\epsilon)
\gamma,\label{eq:mappingequation1}\\
\chi_{n+1}=\chi_n+\mbox{sgn}(\epsilon)\tilde{\rho}_{n+1},
\label{eq:mappingequation2}
\end{gather}
\label{eq:mappingequation}
\end{subequations}
where $\theta_{n} = \langle \hat{\theta}_{n}\rangle$ and
$\tilde{\rho}_{n}=\langle \hat{\rho}_{n} \rangle +
\mbox{sgn}[\pi\ell + \kbar\beta - \gamma(n-1/2)]$. Quantum
accelerator modes are one-to-one related to stable periodic orbits
of this map \cite{Fishman2002,Schlunk2003a,Schlunk2003b}. It is
very important to note that $\epsilon\rightarrow 0$ coincides with
$\hbar\rightarrow 0$ only if $\ell=0$. Otherwise, as in the
experiments here, the classical-particle-like behavior of QAM is
due to a quantum resonance effect.

The stable periodic orbits yielded by Eq.\
(\ref{eq:mappingequation}) (and hence QAM) are classified by their
order $\mathfrak p$ and jumping index $\mathfrak j$ (the number of
momentum units, in terms of the size of the phase-space cell,
traversed after $\mathfrak p$ iterations). The sign of $\mathfrak
j$ is determined by whether this is in the positive or negative
momentum direction. A necessary condition \cite{Fishman2002} for
the existence of a periodic orbit is $ |{\mathfrak j}/{\mathfrak
p}+\mbox{sgn}(\epsilon)\gamma/2\pi|\leq \tilde{k}/2\pi $, which
can be rewritten (for small $\epsilon$) as
\begin{equation}
-|\epsilon| \left( \frac{\phi_{d}}{2\pi} +
\frac{2\ell\gamma}{\kbar^{2}} \right) \leq \frac{\mathfrak
j}{\mathfrak p}+\mbox{sgn}(\epsilon)2\pi
\ell^2\frac{\gamma}{\kbar^{2}} \leq |\epsilon| \left(
\frac{\phi_{d}}{2\pi} - \frac{2\ell\gamma}{\kbar^{2}} \right).
\label{eq:rewrite}
\end{equation}
Both $\phi_{d}$ and $\gamma/\kbar^{2}=gm^{2}/\hbar^{2}G^{3}$ are
independent of $T$, and therefore of $\epsilon$. Equation
(\ref{eq:rewrite}) is convenient when $T$ is varied from just
below to just above $\ell T_{1/2}$, i.e., scanning $\epsilon$ from
negative to positive, as in the experiments described here. As
$\epsilon\rightarrow 0$, the QAM that occur must be characterized
by ${\mathfrak j}$ and ${\mathfrak p}$ such that ${\mathfrak
j}/{\mathfrak p} \rightarrow
-\mbox{sgn}(\epsilon)2\pi\ell^{2}\gamma/\kbar^{2}$. In general
$2\pi\ell^2 \gamma/\kbar^{2}$ is an irrational value, and one
usually observes a succession of increasingly high-order QAM as
$T\rightarrow \ell T_{1/2}$ \cite{Schlunk2003b}. If we tune $g$ so
that $2\pi\ell^{2}\gamma/\kbar^{2}=r/s$, where $r$ and $s$ are
integers, then $\mathfrak j/\mathfrak
p+\mbox{sgn}(\epsilon)2\pi\ell^2\gamma/\kbar^2=0$ for $\mathfrak
j/\mathfrak p=-\mbox{sgn}(\epsilon)r/s$. Once the $({\mathfrak
p},{\mathfrak j})$ QAM satisfying this condition appears, shifting
$T$ closer to $\ell T_{1/2}$ does not result in higher-order
QAM.

\begin{figure}[tbp]
\centering
\includegraphics[width=8.5cm]{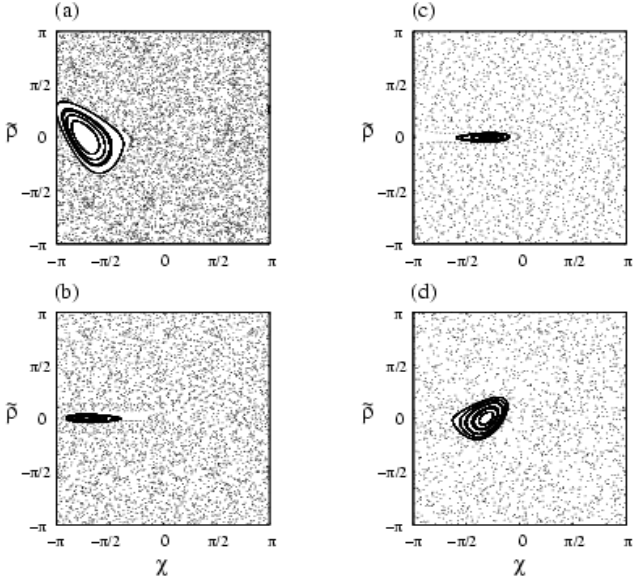}
\caption{Phase space plots produced by Eq.\
(\ref{eq:mappingequation}), when $2\pi \gamma/\kbar^{2}=1/1
\Rightarrow \gamma = (2\pi+\epsilon)^{2}/2\pi$, and
$\tilde{k}=|\epsilon|0.8\pi$ for (a) $\epsilon=-0.88$, (b)
$\epsilon=-0.02$, (c) $\epsilon=0.03$, (d) $\epsilon=0.6$. This
corresponds to $T=57.4$\thinspace $\mu$s, $66.5$\thinspace $\mu$s,
$67$\thinspace $\mu$s, $73$\thinspace $\mu$s. For (a), (b) the
island corresponds to a $({\mathfrak p},{\mathfrak j})=(1,1)$ QAM,
and for (c), (d), to a $({\mathfrak p},{\mathfrak j})=(1,-1)$
QAM.} \label{fig:simpleorder}
\end{figure}

\begin{widetext}

\begin{figure}[tbp]
\centering
\includegraphics[width=8.8cm]{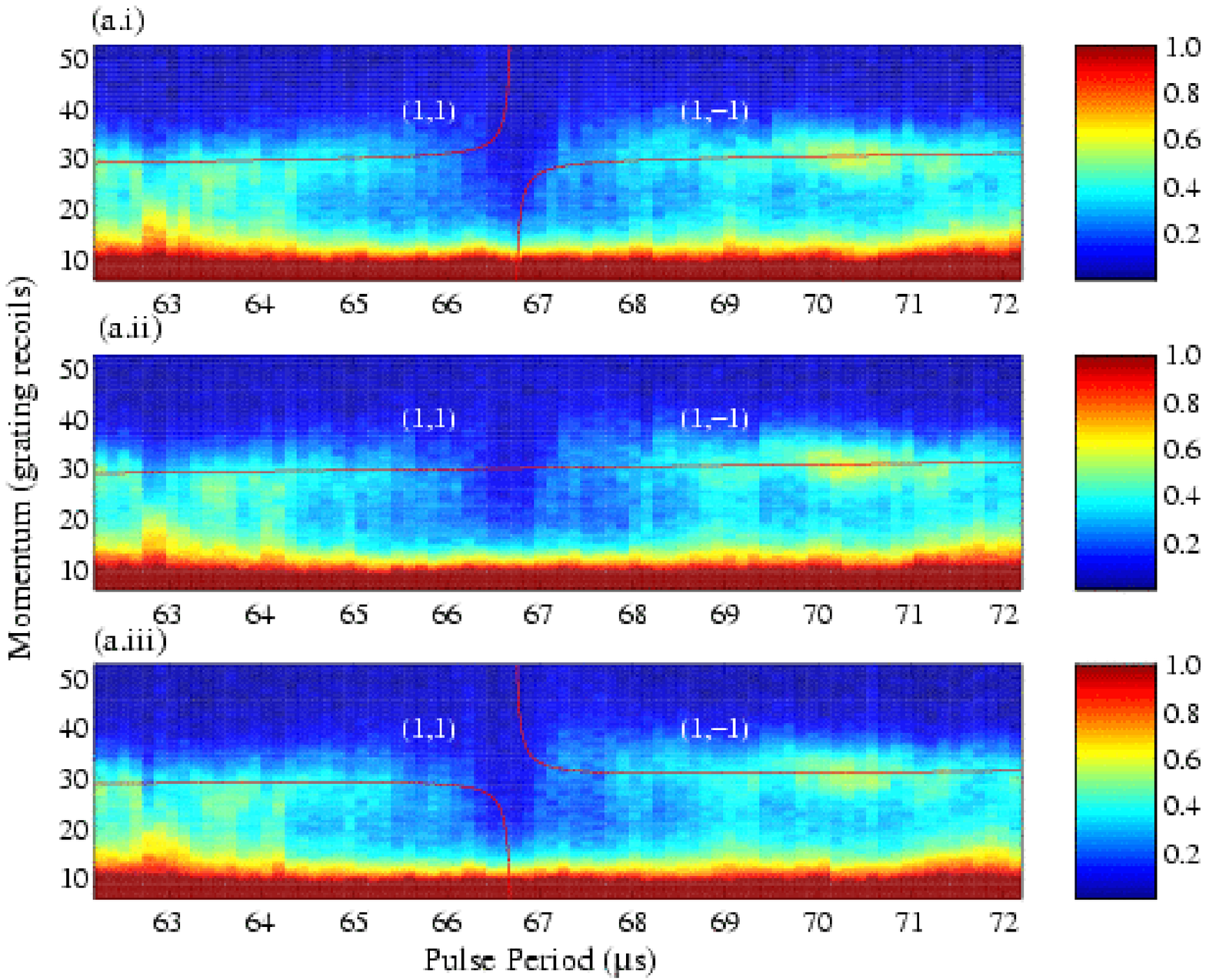}
\includegraphics[width=8.8cm]{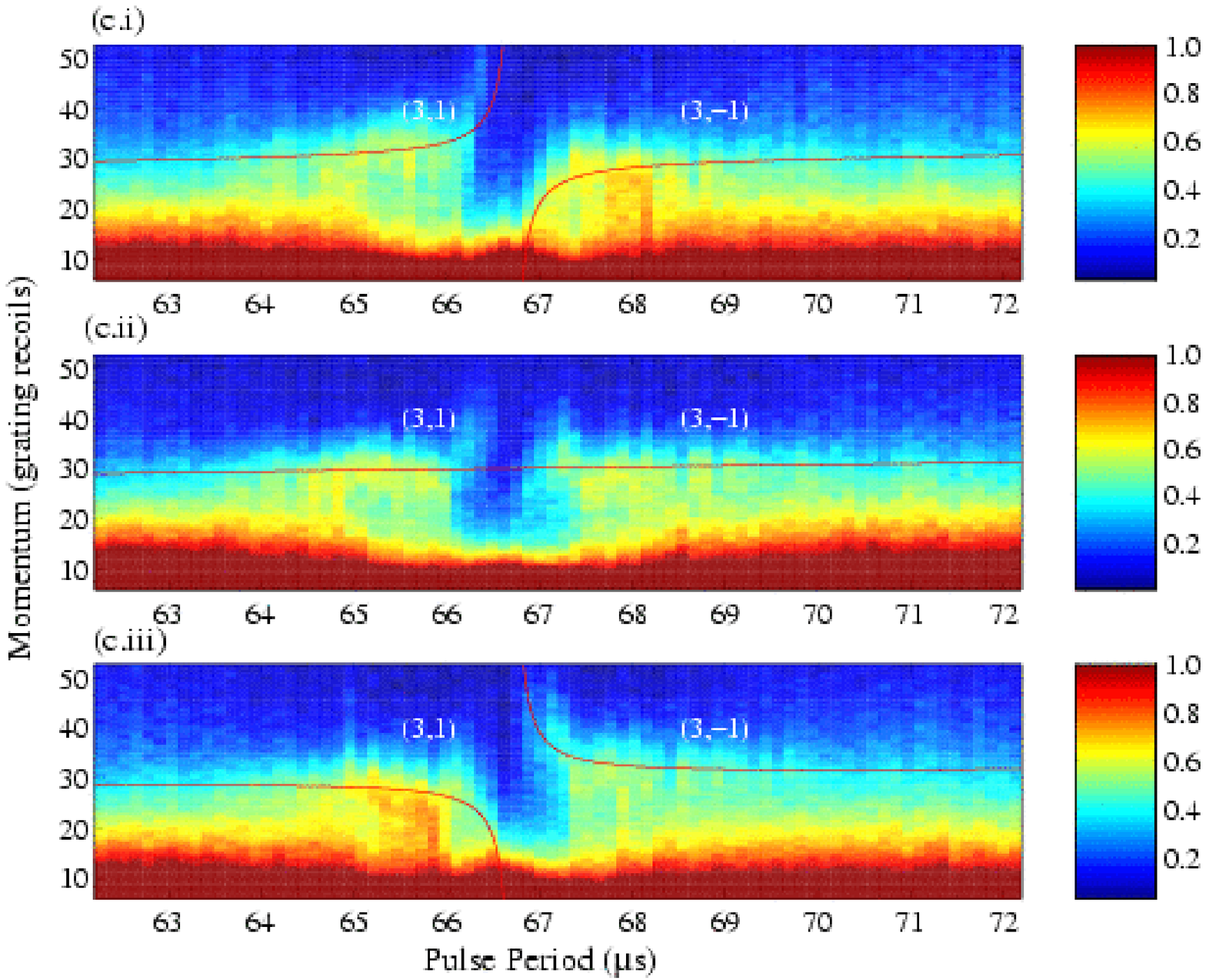}
\includegraphics[width=8.8cm]{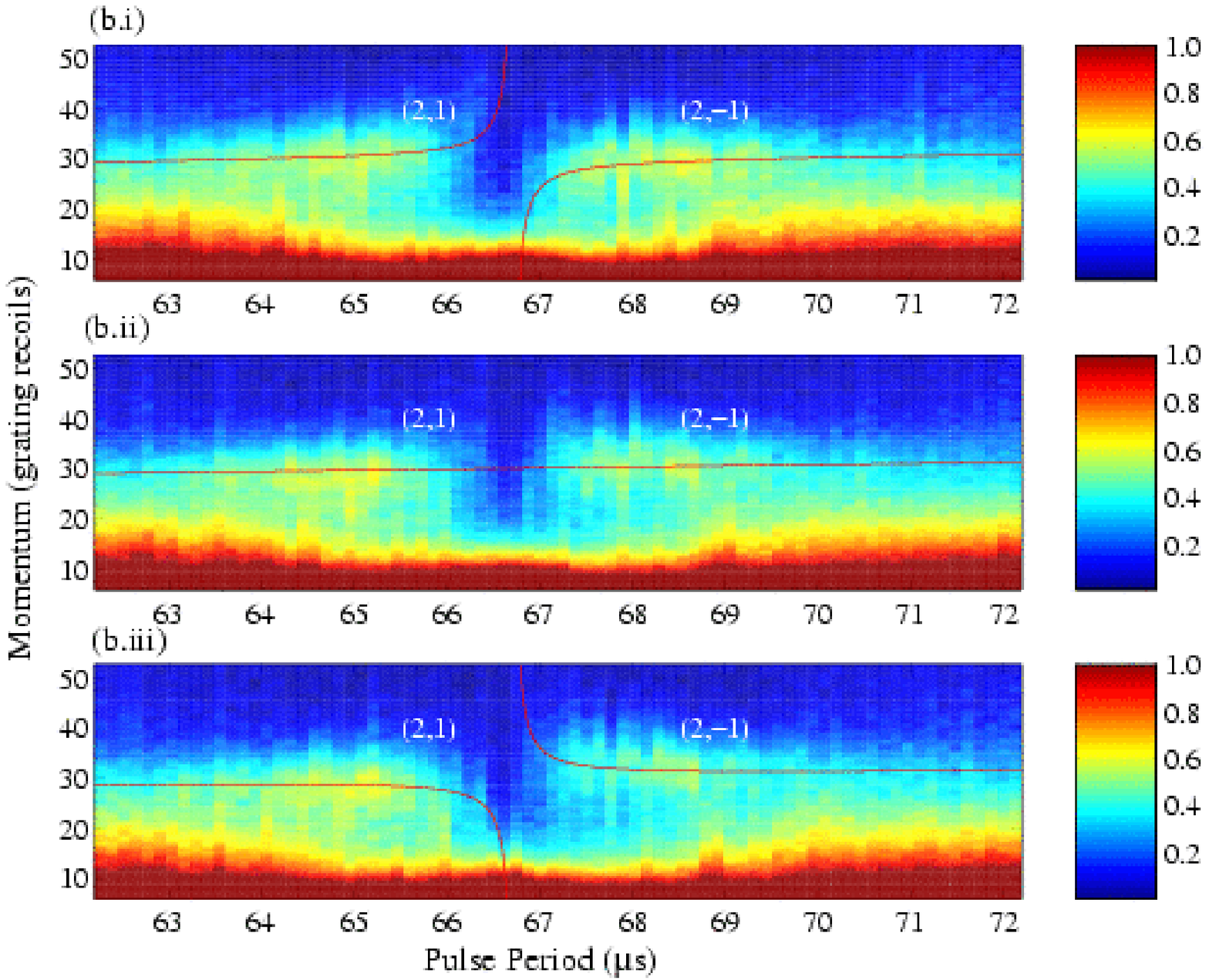}
\includegraphics[width=8.8cm]{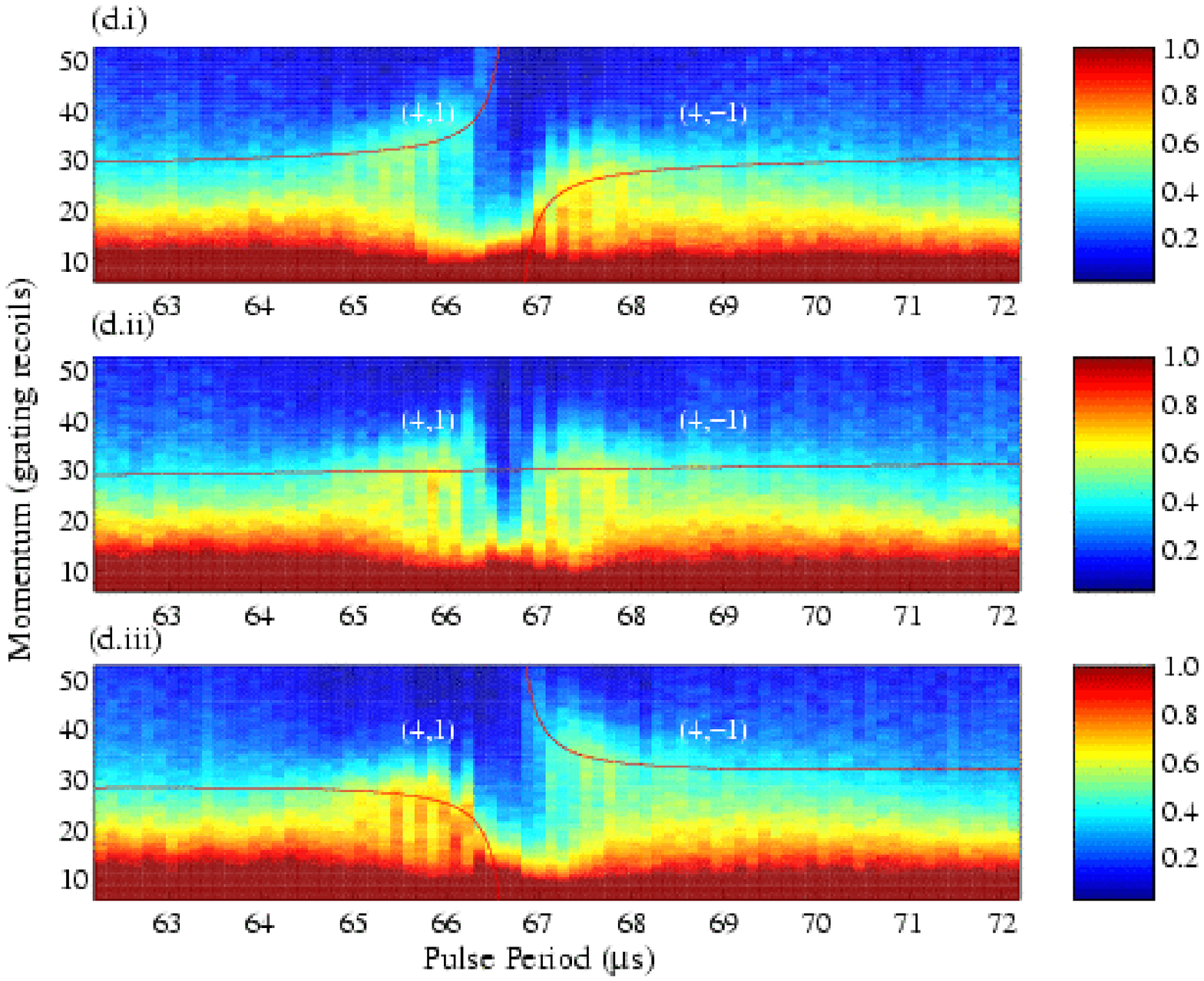}
\caption{(color online). Color density plots of experimental
momentum distributions for different effective gravity $g$
corresponding to (a) $r/s=1/1$ (after 15 kicks), (b) $r/s=1/2$ (30
kicks), (c) $r/s=1/3$ (45 kicks), and (d) $r/s=1/4$ (60 kicks), as
$T$ is varied in the vicinity of the half Talbot time
$T_{1/2}=66.7$\thinspace $\mu$s, from $60.5$\thinspace $\mu$s to
$74.5$\thinspace $\mu$s in steps of $0.128$\thinspace $\mu$s. In
each case the QAM corresponds to ${\mathfrak j}/{\mathfrak
p}=r/s$; subplot (i) corresponds to $w\simeq-8.5\times 10^{-4}$
(deviation from resonant $g$ is $\sim- 8.6\times
10^{-2}$\thinspace ms$^{-2}$), subplot (ii) to $w\simeq 0$, and
subplot (iii) to $w \simeq 8.5\times 10^{-4}$ (deviation from
resonant $g$ is $\sim 8.6\times 10^{-2}$\thinspace ms$^{-2}$).
Overlaid lines, labeled (${\mathfrak p},{\mathfrak j}$), indicate
QAM momenta predicted by Eq.\ (\ref{eq:gravresq}). Population
arbitrarily normalized to maximum value $=1$, and momentum defined
in a frame falling with $g$. Note the significantly greater
population at high momentum (up to $50\hbar G$) near $T_{1/2}$ in
(d.i) and (d.iii), compared to (a.i) and (a.iii).}
\label{fig:linfigs1} \label{fig:linfigs2}
\end{figure}

\end{widetext}

In a frame accelerating with $g$, the momentum after $N$ kicks,
for an initial condition near a ($\mathfrak p, \mathfrak j$)
stable periodic orbit \cite{Fishman2002}, in `grating recoils'
$\hbar G$ \cite{Godun2000} is
\begin{equation}
q_{N}\simeq q_0+N\frac{2\pi}{|\epsilon|}\left[\frac{\mathfrak
j}{\mathfrak p}+\mbox{sgn}(\epsilon)\frac{\gamma}{2\pi}\right],
\label{mom}
\end{equation}
where $q_{0}$ is the initial momentum. For $N$ a multiple of $\mathfrak{j}$, 
this result is exact for
$\epsilon$-classical initial conditions located on
$(\mathfrak{p},\mathfrak{j})$ periodic orbits. We now consider the
momentum of orbits specified by ${\mathfrak j}/{\mathfrak p}=r/s$
(for $\epsilon<0$) and ${\mathfrak j}/{\mathfrak p}=-r/s$ (for
$\epsilon>0$) as a single function of $N$ and $\epsilon$, in the
case where $2\pi\ell^{2}\gamma/\kbar^{2}$ approaches rational
values. Letting $2\pi\ell^2 \gamma/\kbar^{2}=r/s + w \ell^2$
\cite{note1}, we find:
\begin{equation}
q_{N} \simeq q_0 + N\frac{r}{s}\left(
\frac{2}{\ell}+\frac{\epsilon}{2\pi\ell^2}\right) +Nw \left(
\frac{2\pi\ell^2}{\epsilon} + 2\ell + \frac{\epsilon} {2\pi}
\right). \label{eq:gravresq}
\end{equation}
Scanning through $\epsilon$ from negative to positive values, one
does not generally observe two QAM of the same $\mathfrak p$ and
magnitude of $\mathfrak j$ (with positive sign for negative
$\epsilon$, and negative sign for positive $\epsilon$)
\cite{Schlunk2003b,Note:Confusion}. However, in the
gravity-resonant cases we consider, when $2\pi\ell^2
\gamma/\kbar^{2}$ is close to $r/s$, we always observe an $r/s$
and then a $-r/s$ QAM as we scan $\epsilon$ in this way. This is
shown in Fig.\ \ref{fig:simpleorder}, where we plot Poincar\'{e}
sections produced by Eq.\ (\ref{eq:mappingequation}) for
$\gamma=\kbar^{2}/2\pi=(2\pi+\epsilon)^{2}/2\pi$ (i.e., $r/s=1$)
and $\tilde{k}=|\epsilon|\phi_{d} = |\epsilon|0.8\pi$ (the
approximate experimental mean value \cite{dArcy2001a}). The
islands around the $({\mathfrak p},{\mathfrak
j})=(1,-\mbox{sgn}(\epsilon)1)$ periodic orbits remain large over
a wide range of $\epsilon$ and, in dramatic contrast to Ref.\
\cite{Schlunk2003b}, no higher-order island structures appear as
$\epsilon\rightarrow 0$. In Fig.\ \ref{fig:linfigs1}(a) the
corresponding QAM are similarly robust and uninterrupted by
higher-order QAM as $T\rightarrow T_{1/2}$.

From Eq.\ (\ref{eq:gravresq}) we thus see that for a given $N$,
$q$ is a linear function of $\epsilon$ whenever $w=0$. If $w\neq
0$ this changes to a hyperbolic function of $\epsilon$, where the
arms of the hyperbolae point in opposite directions for oppositely
signed $w$.  Deviation from straight line behavior in a QAM
accelerated to a given momentum will be greater for a
gravity-resonant mode corresponding to a smaller value of
$\mathfrak j / \mathfrak p = r/s$. This is because the
acceleration of the mode is $\propto \mathfrak j /\mathfrak p$,
but the deviation is $\propto N$. We consider only QAM where
$\mathfrak j = r = 1$, so high-order modes exhibit, for a given
momentum transfer, greater sensitivity to variations in $g$ than
low-order modes.

In our realization of the quantum $\delta$-kicked accelerator,
$\sim 10^7$ cesium atoms are trapped and cooled in a MOT to a
temperature of $5\mu$K, yielding a Gaussian momentum distribution
with FWHM $6\hbar$G. The atoms are then released and exposed to a
sequence of equally spaced pulses from a standing wave of higher
intensity light $15$\thinspace GHz red-detuned from the
$6^{2}S_{1/2} \rightarrow 6^{2}P_{1/2}$, ($F=4\rightarrow F'=3$)
D1 transition. Hence the spatial period of the standing wave is
$\lambda_{\mbox{\scriptsize spat}}=447$\thinspace nm, and
$T_{1/2}=66.7$\thinspace $\mu$s. The peak
intensity in the standing wave is $\simeq 5 \times 10^4$\thinspace
mW/cm$^2$, and the pulse duration is $t_{p}=500$\thinspace ns.
This is sufficiently short that the atoms are in the Raman-Nath
regime and hence each pulse is a good approximation to a
$\delta$-function kick. The potential depth is quantified by
$\phi_d=\Omega^2 t_p/8\delta_L$, where $\Omega$ is the Rabi
frequency and $\delta_{L}$ the detuning from the D1 transition.
During the pulse sequence, 
\noindent a voltage-controlled crystal phase
modulator is used to stroboscopically accelerate the standing wave
profile. The atoms therefore effectively experience a
non-standard, and controllable, value of gravity. After the
pulsing sequence, the atoms fall through a sheet of laser light
resonant with the $6^{2}S_{1/2}\rightarrow 6^{2}P_{3/2}$,
$(F=4\rightarrow F''=5)$ D2 transition, $0.5$\thinspace m below
the MOT. By monitoring the absorption, the atoms' momentum
distribution is then measured by a time-of-flight method, with
resolution $\hbar G$. For further details see Refs.\
\cite{dArcy2001a,Godun2000}.

In Fig.\ \ref{fig:linfigs1} we show momentum distributions for
experiments in which the value of $T$ was scanned around $T_{1/2}$
($\ell=1$) from $60.5\mu$s to $74.5\mu$s, with
$2\pi\gamma/\kbar^{2}$ varied in the vicinity of $r/s$ equal to
(a) $1/1$, (b) $1/2$, (c) $1/3$, and (d) $1/4$. To maintain the
ideal ($w=0$) total momentum transfer, $15$, $30$, $45$ and $60$
kicks were applied, respectively, fixing $Nr/s$. For each of
Figs.\ \ref{fig:linfigs1}(a), \ref{fig:linfigs1}(b),
\ref{fig:linfigs1}(c), and \ref{fig:linfigs1}(d) the data
displayed are: in subplot (ii), from experiments in which
$2\pi\gamma/\kbar^{2}=r/s$ is fulfilled as exactly as feasible,
yielding linear variation of the QAM momentum with $T$; and in
subplots (i) and (iii), for equal positive and negative
deviations, respectively, from this near-ideality, yielding
hyperbolic variation of the QAM momentum. Typically $\sim$ 10--20
\% of the atoms are accelerated away from the cloud centered at
$p=0$.

In each subplot (ii) of Fig.\ \ref{fig:linfigs1}, the QAM momentum
predicted by Eq.\ (\ref{eq:gravresq}), shown as an overlaid line,
is identical. The expected linear dependence on $T$ appears to be
well confirmed by the data, although the separation of the QAM
from the main, non-accelerated cloud, centered at zero momentum,
is clearer for smaller $s=\mathfrak{p}$ (there is less momentum
diffusion of the main cloud due to the smaller number of kicks).
The effect of imperfectly resonant gravity, shown in subplots (i)
and (iii) in Fig.\ \ref{fig:linfigs1}, is much more dramatic for
larger $s=\mathfrak{p}$, for which more kicks are applied. In
Fig.\ \ref{fig:linfigs1}(a), subplots (i) and (iii) are barely
distinguishable from subplot (ii), whereas in Fig.\
\ref{fig:linfigs1}(d), the momentum distributions in subplots (i)
and (iii) are highly asymmetric compared with subplot (ii), with,
close to $T_{1/2}$, noticeable population at up to $50\hbar G$.
The asymmetry inverts as one changes from below [subplot (i)] to
above [subplot (iii)] the resonant value of gravity. We therefore
observe a clear qualitative change in the QAM dynamics, highly
sensitive to a control parameter. The displayed predictions of
Eq.\ (\ref{eq:gravresq}) show that deviations from linear behavior
only occur when very close to $T_{1/2}$ in Figs.\
\ref{fig:linfigs1}(a.i) and \ref{fig:linfigs1}(a.iii), but are
much more significant in Figs.\ \ref{fig:linfigs1}(d.i) and
\ref{fig:linfigs1}(d.iii). This is due to the larger number of
kicks necessary for large $s=\mathfrak{p}$ to achieve the same QAM
momentum.

The procedure of determining the `standing wave acceleration' at
which straight-line behavior of a given $({\mathfrak p},{\mathfrak
j})$ QAM momentum is observed as a function of $T$ could, in
principle, be used as a sensitive atom-optical means of relating
$h/m$ \cite{Gupta2002} to the local gravitational acceleration
\cite{Peters1999}. This is because
$2\pi\gamma/\kbar^{2}=r/s$ can be rephrased as
$g=(h/m)^{2}(r/s)/\lambda_{\mbox{\scriptsize spat}}^{3}$ and would
be determined by noting when the {\it total}\/ acceleration
(sinusoidal potential plus gravitational) causes these equalities
to be fulfilled for a known $r/s$, and then subtracting the
imposed acceleration of the potential. In our setup, where the
sinusoidal potential is `accelerated' by using a crystal phase
modulator to phase-shift the retroreflected laser beam
\cite{dArcy2001a,Godun2000}, the value of the phase shift due to a
particular applied voltage is difficult to calibrate more
precisely than $\sim 1$\thinspace \%. This accordingly limits our
measured precision of the relationship between the local
gravitational acceleration and $h/m$ to $\sim 1$\thinspace \%.
Accurate prediction of the QAM momenta for imperfectly resonant
values of the effective gravity, as displayed in Fig.\
\ref{fig:linfigs1}, is also hampered. This could be improved by a
configuration in which a moving sinusoidal potential is formed by
two counterpropagating beams with a controllable frequency
difference \cite{Denschlag2002}, where calibration of the phase
shift to between $1$ppm and $1$ppb is possible. Calibration of
$\lambda_{\mbox{\scriptsize spat}}$ to less than $1$ppb is also
feasible \cite{Peters1999}, allowing for the possible sensitive
determination of either the local gravitational acceleration
\cite{Peters1999} or $h/m$ \cite{Gupta2002}, depending on which is
known more precisely at the outset. The feasibility of any such
scheme will ultimately depend on how precisely the atomic
ensemble's dynamics permit the determination of the acceleration
of the sinusoidal potential for which the resonant, linear with
$T$, behavior of the QAM occurs. Ascertaining this will require
substantial theoretical and experimental investigation.

In conclusion, we have observed qualitative changes in the
motional quantum dynamics of cold cesium atoms, which are highly
sensitive to the precise value of an externally adjustable
parameter, the effective gravity. This is distinct from
conceptually related proposals that consider slightly differing
Hamiltonians to study the Loschmidt echo or fidelity, and
demonstrates an attractive link to the concepts of highly
sensitive dynamics in classically chaotic systems. Furthermore, we
have described a feasible experimental scheme taking advantage of
this sensitivity to determine a relationship between the local
gravitational acceleration and $h/m$.

We thank K. Burnett, S. Fishman, I. Guarneri, L. Rebuzzini, G.S.
Summy, and particularly R.M. Godun, for very helpful discussions.
We acknowledge support from the Clarendon Bursary, the UK EPSRC,
the Royal Society, the EU through the TMR `Cold Quantum Gases'
Network, the Lindemann Trust, and NASA.


\begin{thebibliography}{99}
\bibitem{note1}
We have multiplied out a singularity occurring for $q$ at
$\epsilon=0$  when $w=0$.
\bibitem{Note:Confusion}
The sign of ${\mathfrak j}$ used here is consistent with that used
for the gravitional potential, and the definition of ${\mathfrak
j}$ in Ref.\ \cite{Fishman2002}.
\end{thebibliography}
\end{document}